%
%
% Last modified by M.W., 20/11/96. This version submitted to
% Chaos, Solitons and Fractals, 21/11/96.
%
% Page break inserted at start of section 4.
%
%
\def\indentoff{\parindent=0pt}
\def\indenton{\parindent=20pt}
\def\gap{\vskip 6pt}
\def\ket#1{\vert #1\rangle }
\def\bra#1{\langle #1\vert }

\def\ref#1#2#3#4#5#6{[#1] #2, {\sl #3}, {\bf #4}, #5, (#6).\gap}
\magnification=\magstep1
\vsize=25 true cm
\goodbreak
\indenton
\vskip 1cm
\centerline{\bf DISTRIBUTION OF OSCILLATOR STRENGTHS FOR}
\centerline{\bf RECOMBINATION OF LOCALISED EXCITONS IN TWO DIMENSIONS} 
\vskip 1cm
\centerline {Michael Wilkinson, Paul N. Walker and Kaumba Chinyama}
\vskip 1cm
\centerline {Department of Physics and Applied Physics,}
\centerline {John Anderson Building,}
\centerline {University of Strathclyde,}
\centerline {Glasgow, G4 0NG,}
\centerline {Scotland, U.K.}
\vskip 2cm
\centerline{\bf Abstract}
\gap
We investigate the distribution of oscillator
strengths for the recombination of excitons in
a two dimensional sample, trapped in local minima of the
confinement potential: the results are derived
from a statistical topographic model of the potential.
The predicted distribution of oscillator strengths is
very different from the Porter-Thomas distribution
which usually characterises disordered systems,
and is notable for the fact that small oscillator
strengths are extremely rare.
\vfill
\eject
%
%
%
%%%%%%%%%%%%%%%%%%%%%%%%%%%%%%%%%%%%%%%%%%%%%%%%%%%%%%%%%%%%%%%%%%%%%
%
\noindent {\bf 1. Introduction}
\gap
Many experiments have been performed in which excitons
(excitations in the form of an electron-hole bound state) 
are trapped in a layer of lower bandgap semiconductor, between
two layers of higher bandgap material [1,2]. Exciton absorption 
and emission lines in these two-dimensional samples are 
typically much broader than in three-dimensional systems.
This is often due to fluctuations in the width of the 
layer: there is a \lq quantum confinement' contribution to 
the energy of the exciton, which is a decreasing function $E(w)$
of the layer width $w$, and which is analogous to the
ground state energy $\pi^2\hbar^2/2mw^2$ of a particle
trapped in a one-dimensional potential well [3,4].
 
The excitons are able to interact with phonons by 
radiationless processes [5,6], becoming trapped in local minima
of the confinement energy: the lifetime for electron-hole
recombination is typically long enough that most of the luminescence
observed from these samples is from trapped excitons.
This trapping effect has recently been observed directly
by spatially resolved studies of exciton luminescence [7,8]:
if the luminescence is recorded from a macroscopic area 
of the sample, a broad spectrum is observed, whereas
sufficiently small microscopic areas show either
no luminescence, or a small number of relatively sharp lines,
corresponding to excitons in a single trap.
These exciton traps may be thought of a novel type of 
mesoscopic system.

The spectra of two dimensional excitons show some near-universal
features, for example it was noticed that the Stokes shift $S$ of 
the luminescence spectrum relative to the absorption peak, and 
the width $W$ of the absorption spectrum usually satisfy 
$S/W\approx 0.6$, independent of the semiconductor materials
or of the magnitude of the broadening [9]. This observation was 
explained using a classical picture of trapping of excitons 
in local minima of a smooth effective confinement potential,
modelled as a Gaussian random function [10]. This statistical
topographic model predicted $S/W=0.55..$, in good agreement with 
the experimental values. In section 2 we review this model,
and discuss a variant of the experimental approach which
may give better agreement with theoretical predictions.

In this paper we discuss the distribution of oscillator
strengths (or, equivalently, Einstein A coefficients) 
for exciton recombination, for excitons localised in
minima of the effective confinement potential. We use the 
same statistical topographic model as was previously used
to calculate the Stokes shift. If the model is correct, 
the probability distribution of oscillator strengths for
localised excitons should be a near universal signature
of trapped excitons, but we believe that this may be
a more sensitive test of the validity of the model.
The predicted probability distribution of oscillator 
strengths is 
$$P(I)=I^{-9}\exp(1/2I^4){\rm erfc}(\sqrt{3}/2I^2)\eqno(1.1)$$
(the scaling of the intensities $I$ is arbitrary; for this
form of the distribution the mean intensity is $\langle I\rangle
\approx 0.699..$). 
This result is interesting because the predicted distribution 
has very few small intensities: the fraction of recombination 
lines predicted to be less than half the mean intensity is 
approximately $3.15\times 10^{-7}$.
It is very different from the Porter-Thomas distribution,
$P(I)\sim \exp(-I)/\sqrt {I}$, which usually characterises the
distribution of oscillator strengths for disordered
or complex systems [11], and which follows from the Gaussian distribution
of matrix elements in such systems.

Section 3 discusses the model used for calculating
the matrix elements for recombination of excitons 
trapped in local mimima of a potential. In section 4
we give a complete characterisation of the distribution of 
quadratic forms characterising the stationary points of 
an isotropic Gaussian random function: although this is not 
difficult to obtain using standard techniques of statistical
topography [12,13], we could not find a discussion of it in 
the literature in the form we require. 
These results are then used to obtain (1.1).

The field of statistical topography was 
largely stimulated by the desire to understand optical
properties of random surfaces, such as that of the sea.
In section 4 we also comment on the closest optical analogy of
our results, namely the distribution of intensities of
reflected images of a small light source on a distant
random surface (for example, \lq sea glitter', reflections 
of the sun from the sea, observed from an aircraft [14]).
The distribution (1.1) turns out to be very different from
that of the intensities of sea glitter sparkles.
\gap\gap\gap
%
%
%
%%%%%%%%%%%%%%%%%%%%%%%%%%%%%%%%%%%%%%%%%%%%%%%%%%%%%%%%%%%%%%%%%%%%%
%
\noindent {\bf 2. The statistical topographic model}
\gap
Here we briefly review the statistical topographic model
for exciton luminescence presented in [9,10]. We also propose a 
reason for the small discrepancy between theory and experiment,
and a variant of the experimental approach which may give a 
better agreement with experiment.

Figure 1 shows the absorption (a) and luminescence (b) spectra
of excitons in a semiconductor heterostructure: the data
are taken from [15]. First consider the form of the absorption peak. 
In the absence of disorder the peak would be very sharp, because 
(unlike unbound electron-hole pairs) conservation of momentum implies 
that the oscillator strength for creation of the exciton vanishes
unless the centre of mass of the exciton is stationary ([16]; this 
can also be verified using the model discussed in section 3). 
The width of the absorption peak in 
figure 1(a) is determined by inhomogeneous broadening 
due to disorder. The energy of an exciton in the ground state depends 
upon the width of the potential well in which it is confined, and the 
well width varies randomly with position in the plane. We assume that 
the length scale over which the well width varies is large compared 
to the width of the wells: this implies that the ground state energy 
of a static exciton at position $(x,y)$ is a well defined smooth function, 
which we denote by $E(x,y)$. It is reasonable to assume that the 
fluctuations of $E(x,y)$ represent contributions from many independent
events, and the central limit concept then indicates that the 
fluctuations are Gaussian. Because the excitons are created
with zero centre of mass motion, the absorption spectrum
is proportional to the distribution of $E(x,y)$. The
curve (c) in figure 1 is a fit of a Gaussian curve to
the absorption spectrum (a): it fits quite closely.

Now consider the luminescence peak. This is shifted toward lower energies
because the exitons can lose energy before they decay. Time-resolved studies
of spectral hole burning indicate that the energies of excitons can change
over a timescale of typically tens of picoseconds, much shorter than the 
half-life for decay of excitons, typically several hundred picoseconds [5,6]. 
The predominant mechanism 
of energy loss for the excitons appears to be by the excitation 
of phonons: if the absorption spectrum is probed with 
narrow spectral lines, it is possible to observe features in the 
luminescence spectrum which are shifted from the probe frequency by 
multiples of the frequency of the optical phonons [6]. These 
results justify the following picture of the luminescence 
process: after the exciton is created at position $(x_0,y_0)$ with energy 
$E_0=E(x_0,y_0)$, it will move into regions where the potential energy 
$E(x,y)$ is less than $E_0$, and the excess energy $E_0-E(x,y)$ appears 
as kinetic energy. The moving exciton is able to excite
phonons, and as it does so it loses kinetic energy. Eventually, if it does
not decay in the meantime, it will end up trapped in a local minimum of the 
potential energy $E(x,y)$. Because the exciton lifetime 
is much longer than the timescale associated with energy transfer to 
phonons, most of the excitons are trapped close to a local minimum of 
the potential energy $E(x,y)$ when they decay. 

These considerations lead to a model in which both the
absorption and luminescence spectra are determined by
the statistical topography of a Gauss random function, which
we will denote by $f(x,y)$, and which represents the energy
function $E(x,y)$ after applying linear scaling transformations 
to $E$, $x$ and $y$ such that 
$$\langle f\rangle=0,\ \ \ \langle f^2\rangle=1$$
$$\langle f_x^2\rangle=\langle f_y^2\rangle=1
\ .\eqno(2.1)$$
The absorption spectrum is proportional to the
Gaussian probability distribution of this function, whereas
the luminescence spectrum is determined by the distribution
of heights of its local minima.

The Gaussian random function $f(x,y)$ is characterised completely
by its correlation function, $C({\bf r})$, which in the isotropic
case is a function of $r=|{\bf r}|$ only. The calculations
in [13], [10] show that the distribution of stationary points 
depends on the correlation function only through averages of second 
derivatives. In the isotropic case, after scaling the function
so that (2.1) is satisfied, the distribution of stationary
points is characterised by a single parameter $a$ [10]:
$$\langle f_{xx}^2\rangle=\langle f_{yy}^2\rangle
=3\langle f_{xy}^2\rangle=3a
\ .\eqno(2.2)$$
The distribution of heights of local minima was determined analytically in
[9,10]: the result is
$$P_{\rm min}(f)={\sqrt{3}\over {2\pi a\sqrt{2a-1}}}\biggl[
-(2a-1)f\exp\biggl({-af^2 \over{(2a-1)}}\biggr)$$
$$+{a\sqrt{2\pi a(2a-1)}\over{\sqrt{3a-1}}}
\exp\biggl({-3af^2\over{2(3a-1)}}\biggr)
{\rm erfc}\biggl(\sqrt{a\over{2(2a-1)(3a-1)}}f\biggr)$$
$$+{\textstyle{1\over 2}}\sqrt{2\pi (2a-1)}(f^2-1)
\exp\bigl(-f^2/2\bigr)
{\rm erfc}\biggl({f\over {\sqrt{2(2a-1)}}}\biggr)
\biggr]\eqno(2.3)$$
where ${\rm erfc}(x)$ is the complementary error function [17]. 

The prediction (2.3) contains the undetermined parameter
$a$. In [9,10] it was argued that the annealing process involved in
the growth of the heterostructures causes the fluctuations
of $E(x,y)$ to be suppressed by a diffusive process: we
write
$$E(x,y)=\int dx'\int dy' P(x-x',y-y')\,E'(x',y')$$
$$P(X,Y)={1\over {8\pi Dt}}\exp[-(X^2+Y^2)/4Dt]\eqno(2.4)$$
where $E'(x,y)$ characterises an initial distribution of well width
fluctuations with a much shorter correlation length,
$D$ is the diffusion constant and $t$ the annealing time. 
We may therefore model $E(x,y)$ as the convolution of a white 
noise function with a Gaussian. It follows that the correlation 
function of $E(x,y)$ is also a Gaussian, implying that $a=1$.

The distribution of minima (2.3) with $a=1$ is shown as 
curve (d) in figure 1: the mean and variance have been 
scaled to correspond to the mean and variance of the 
Gaussian distribution (c). The curve is not a particularly
good fit to the luminescence spectrum, (b).
The experimental data shown in figure 1 are typical: it
is usually found that the Stokes shift is somewhat higher
than our theoretical prediction [9]. We will make two points about
this discrepancy.

First, we propose a qualitative explanation of this observation:
the excitons can move into deeper local minima than
the ones in which they were initially trapped before they
recombine, either by thermally assisted hopping or quantum mechanical
tunnelling. This explanation is supported by the literature on time 
resolved luminescence studies, which show that the Stokes shift
initially assumes a value close to our prediction, and
then slowly increases [18]. 

Our second, more important, point concerns
resonant Rayleigh scattering experiments on excitons, such as that
discussed in [15]. 
>From the discussion above, it is apparent that it would be desirable to
have a more direct probe of the density of local minima of the effective 
potential, to facilitate comparison between theory and experiment.
We will now argue that resonant Rayleigh scattering spectra measure the
density of local minima of the effective potential. The intensity of 
Rayleigh scattering from an exciton will be proportional to the time
over which the exciton survives at the energy of the incident radiation.
The lifetime of an exciton trapped in a minimum of the effective potential
is equal to the lifetime for luminescent decay, typically less than 
a nanosecond.
An exciton which is not trapped may lose energy by exciting phonons, and
the lifetime for these processes is much shorter, typically a few tens 
of picoseconds. The spectra for resonant Rayleigh scattering are proportional
to a density of states weighted by the exciton survival time. Because the 
survival time is very much larger for trapped excitons, the Rayleigh 
scattering spectrum is weighted very heavily by the density of states for 
trapped excitons. 

The isolated points plotted on figure 1 are resonant Rayleigh 
scattering amplitudes measured at a discrete set of frequencies, taken
from data published in [15]. Both the mean and the variance
of these data are remarkably close to the theoretical distribution of 
local minima, curve (d), which is a plot of (2.3) with $a=1$,
and with the mean and variance chosen to match those of the Gaussian 
fit to the {\sl absorption} spectrum. 
\gap\gap\gap
%
%
%
%%%%%%%%%%%%%%%%%%%%%%%%%%%%%%%%%%%%%%%%%%%%%%%%%%%%%%%%%%%%%%%%%%%%%
%
\noindent {\bf 3. Model for exciton luminescence matrix elements}
\gap
We consider a solid consisting of N atoms, each of which can
exist in four different states:
\gap
\noindent 1. Neutral and unexcited, $\ket {0}$.
\par
\noindent 2. With additional electron (in the conduction band), $\ket {-}$.
\par
\noindent 3. With an electron removed (i.e. with hole in valence band), 
$\ket {+}$.
\par
\noindent 4. Excited atom (or Frenkel exciton), with an electron in the 
conduction band and a hole in the valence band, $\ket {e_F}=\ket {\pm}$.
\gap
We consider a Hilbert space for the solid with $4^N$ basis
vectors, consisting of all possible combinations of these
four states of the $N$ individual atoms: for example
a state with two electrons, a hole and an exciton could
be written $\ket {0,0,-,0,0,..,0,0,+,0,\pm,0,-,0,..}
=\ket {n_1,n_2,n_3;n_3,n_4}$, where $n_1,n_2$ are the
positions of the electrons, $n_4$ is the position of the hole,
and $n_3$ is the position of the Frenkel exciton.
 
This model is reasonable for insulators in which the 
electrons are tightly bound to individual atoms. This
description can also be carried over directly to 
semiconductors, if the electron and hole states
localised on individual atoms are replaced by Wannier
states derived by integrating over the conduction and
valence band wavefunctions respectively [16].

In a semiconductor, the excitons are typically of the
Wannier type, in which the electrons and holes, although
correlated, are not bound to the same atomic orbital.
The Wannier exciton state $\ket {e_W}$ is a superposition of 
electron and hole states of the form
$$\ket {e_W}=\sum_{n_1,n_2} c_{n_1,n_2}\ket {n_1;n_2}\eqno(3.1)$$
where $n_1$, $n_2$ are the postions of the electron and hole
respectively. In the case of the weakly localised Wannier
exciton, the coefficients $c_{n_1,n_2}$ can be approximated
by a continuous wavefunction $\psi ({\bf x},{\bf y})$, where
${\bf x}$ and ${\bf y}$ are the locations of atoms with labels 
$n_1$ and $n_2$. The wavefunction $\psi ({\bf x},{\bf y})$
is an eigenfunction of an effective Hamiltonian [16]
$$\hat H={1\over{2m_e}}{\bf p}_{\rm e}^2+{1\over{2m_h}}{\bf p}_{\rm h}^2
+{e^2\over{4\pi \epsilon_0}}{1\over{|{\bf r}_{\rm e}-{\bf r}_{\rm h}|}}+
V_{\rm e}({\bf r}_{\rm e})+V_{\rm h}({\bf r}_{\rm h})
\ .\eqno(3.2)$$
\par
The intensity of emission from an exciton state is 
proportional to the square of the dipole matrix
element $\bra {e_W}\hat X\ket {0}$ for the transition
between the exciton state and the ground state of the
system, summed over three orthogonal choices for the
coordinate $X$. This matrix element can be expressed in terms 
of the dipole matrix elements of the localised basis
states. We assume that the dipole matrix element for 
the basis states is 
$$\bra {0} \hat X\ket {n_1;n_2}=\epsilon\ \delta_{n_1 n_2}\eqno(3.3)$$
i.e. the dipole matrix element for recombination is assumed
to be negligible, unless the electron and hole are on
the same atomic site. Most crystals have centres of symmetry,
and the dipole matrix element $\bra {0}\hat X\ket {n;n}$ vanishes
if the atomic orbitals or Wannier functions 
defining the valence and conduction bands have the same
parity with respect to the $X$ coordinate.
The atomic orbitals associated with 
successive bands typically have opposite parity with respect
to one of the coordinates, so that for at least one 
choice of the coordinate $X$ the matrix element 
considered in (3.3) does not vanish because of symmetry
considerations. 

The required matrix element can now be calculated using (3.3):
$$\bra {0}\hat X\ket {e_W}
=\sum_{n_1 n_2}c_{n_1 n_2} \bra {0}\hat X \ket {n_1;n_2}$$
$$=\epsilon \sum_n c_{nn}
\sim \epsilon' \int d{\bf r}\ \psi({\bf r},{\bf r})
\eqno(3.4)$$
(here $\epsilon'$ is another constant).
The matrix element is therefore proportional to the amplitude
for the electron and hole to be at the same site. 

We now consider the case of excitons in heterostructures, where 
the excitons are trapped in a layer of low bandgap material 
between regions of higher bandgap. Imperfections of the growth
process result in random fluctuations of the layer width:
we will assume that the fluctuations of the layer width are
on a larger scale than the size of the exciton [10], and we 
therefore use the following model
$$V_{\rm e}({\bf r}_{\rm e})+V_{\rm h}({\bf r}_{\rm h})=
v_{\rm e}(z_{\rm e})+v_{\rm h}(z_{\rm h})+u_{\rm e}({\bf x}_{\rm e})+
u_{\rm h}({\bf x}_{\rm h})$$
$$\sim v_{\rm e}(z_{\rm e})+v_{\rm h}(z_{\rm h})
+u_{\rm e}({\bf X})+u_{\rm h}({\bf X})\eqno(3.5)$$
where ${\bf x}_{\rm e/h}=(x_{\rm e/h},y_{\rm e/h})$ are the electron/hole
coordinates in the $(x,y)$ plane, and ${\bf X}$, ${\bf x}$ are
the corresponding centre of mass and relative coordinates.
In the second line the assumption that the variation of the 
potentials $u_{\rm e/h}$ is slow on the scale of the exciton
diameter justifies the approximation 
${\bf x}_{\rm e}\sim {\bf x}_{\rm h}\sim {\bf X}$. 
The effective Hamiltonian can then be written in the
separated form
$$H({\bf r}_e,{\bf r}_h,{\bf p}_e,{\bf p}_h)=H_0+H_1$$
$$H_0={1\over {2\mu }}{\bf p}^2+{e^2\over{4\pi \epsilon_0|{\bf r}|}}
+v_e(z_e)+v_h(z_h)$$
$$H_1={1\over{2M}}{\bf P}^2+V_{\rm eff}({\bf X})\eqno(3.6)$$
where $V_{\rm eff}(X,Y)=u_{\rm e}(X,Y)+u_{\rm h}(X,Y)$, 
${\bf P}$ is the momentum conjugate to ${\bf X}=(X,Y)$,
and $\mu $, $M$ are the reduced and total masses. 
The corresponding solution of the Schr\" odinger equation is a product of 
an exciton wavefunction $\chi (x,y,z_e,z_h)$, and a wavefunction 
$\phi (X,Y)$ for the centre of mass motion in the $(X,Y)$ plane.
Equations (3.4) and (3.6) show that the transition strength for 
exciton recombination is of the form
$$I\sim |\bra {0}\hat X\ket {e_W}|^2=C \bigg|\int dX \int dY \ 
\phi (X,Y)\bigg|^2
\eqno(3.7)$$
where the constant $C$ is the same for all exciton states.

We first discuss the interpretation of this result for 
a non-disordered system. In this case, the centre of mass 
wavefunctions are $\phi ({\bf X})=A^{-1/2}\exp[i{\bf k}.{\bf X}]$,
where $A$ is the area of the sample. 
The recombination transitions only occur from the ground state, for which
$\phi (X,Y)=A^{-1/2}$, because the integral in (3.7) vanishes for 
all of the other states. Equation (3.7) then implies that the 
transition rate for recombination from the centre
of mass ground state is proportional to the area of the 
system, whereas the rate for all of the other possible states
is zero. This is in accord with the expectation that the 
sum of the transition rates for a system of area $A$
should be proportional to $A$.
 
Now consider the case of a disordered system. We will
use the same model as in [9] and [10]: we assume that
the excitons interact strongly with phonons, and that
mobile excitons rapidly lose energy by 
exciting phonons: most of the exciton recombination 
therefore occurs after the excitons have become trapped in 
minima of the effective potential $V_{\rm eff}(X,Y)$.
We will assume that the effective potential for the centre
of mass motion is an isotropic Gauss random function.
The form of the effective potential in the neighbourhood
of its minima can be approximated by a quadratic form: 
if the origin of the $(X,Y)$ plane is shifted to lie at 
the minimum, we write
$$V_{\rm eff}(X,Y)\sim V_0+{\textstyle{1\over 2}}
[V_{XX}\,X^2+V_{YY}\,Y^2+2V_{XY}\,XY]=
V_0+{\textstyle{1\over 2}}{\bf X}^T\tilde M {\bf X}\eqno(3.8)$$
where $\tilde M$ is the Hessian matrix of second derivatives
evaluated at the minimum. The exciton recombination
occurs from a ground state of the centre of mass motion
trapped in this minimum, for which the wavefunction
$\phi (X,Y)$ is a harmonic oscillator ground state,
which satisfies
$$\int dX \int dY\ \phi(X,Y)=c[{\rm det} \tilde M]^{-1/8}\eqno(3.9)$$
where $c$ is independent of the local environment in which
the exciton is trapped. The distribution of transition
strengths $I$ for trapped excitons is therefore determined from
the distribution of determinants of the Hessian matrix at
minima:
$$I\sim |D|^{-1/4},\ \ \ D={\rm det}(\tilde M)
\ .\eqno(3.10)$$
\gap\gap\gap
\vfill
\eject
%
%
%%%%%%%%%%%%%%%%%%%%%%%%%%%%%%%%%%%%%%%%%%%%%%%%%%%%%%%%%%%%%%%%%%%%%
%
\noindent {\bf 4. Distributions of intensities}
\gap
\noindent {\sl 4.1 Distribution of quadratic forms at statioary points}
\gap
Methods for calculating properties of point singularities
such as minima are well known: the one and two dimensional
cases are discussed by Rice [12] and Longuet-Higgins [13]. 
We will calculate the distribution of the trace $T$
and determinant $D$ of the Hessian matrix 
$$\tilde M=\pmatrix{f_{xx}&f_{xy}\cr
                    f_{xy}&f_{yy}\cr}
\eqno(4.1)$$
describing the second derivatives $f_{xx},f_{xy},f_{yy}$ of an
isotropic Gauss random function $f(x,y)$ at its stationary points.

We follow the approach and notation of [10]: the distribution 
of extrema is determined by the joint probability 
distribution $P(f,f_x,f_y,f_{xx},f_{yy},f_{xy})$ of the 
function $f$, its first derivatives $f_x,f_y$, and its second 
derivatives evaluated at the same point $(x,y)$. By a simple
adaptation of the calculation in [10],
the joint distibution of the trace and determinant is
$$P(T,D)={1\over {\cal N}}\int df \int df_{xx}\int df_{yy}\int df_{yy}
\,P(f,0,0,f_{xx},f_{yy},f_{xy})\,
|d|\,\delta (D-d)\,\delta(T-t)\eqno(4.2)$$
where ${\cal N}$ is a normalisation factor, and $t=(f_{xx}+f_{yy})$,
$d=f_{xx}f_{yy}-f_{xy}^2$. The function
is assumed to be scaled so that (3.1) is satisfied.

In terms of the variables
$R={1\over 2}t$, $X={1\over 2}(f_{xx}-f_{yy})$, $Y=f_{xy}$,
the probability distribution in (4.2) was obtained in [10]:
$$P(f,0,0,f_{xx},f_{yy},f_{xy})={1\over{(2\pi)^3a\sqrt{2a-1}}}\exp(-f^2/2)$$
$$\times\exp[-(f+R)^2/2(2a-1)]\exp[-(X^2+Y^2)/2a]
\ .\eqno(4.3)$$
The integral (4.2) can now be evaluated easily: the normalised
probability density is
$$P(T,D)={1\over{32a^2}}\,\sqrt{{3\over{\pi a}}}\,|D|\,\exp[D/2a]\,
\exp[-3T^2/16a]\,\Theta (D-T^2/4)\eqno(4.4)$$
where $\Theta (x)$ is a step function, decreasing from $1$
to $0$ at $x=0$.
\gap
\noindent {\sl 4.2 Distribution of oscillator strengths}
\gap
Now we will use the results of section 3 to calculate
the distribution of transition strengths for excitons.
The distribution of the determinant $D$ of the Hessian
matrix at stationary points is obtained by integrating (4.4) over $T$:
$$P(D)=\int_{-\infty}^\infty dT\ P(T,D)=
{1\over{8a^2}}\,|D|\,\exp(D/2a)\,
\biggl[\Theta (D)+\Theta (-D)\,{\rm erfc}({\textstyle{1\over 2}}\sqrt{3D/a})
\biggr]
\ .\eqno(4.5)$$
It is noteworthy that the parameter $a$ only appears in the
ratio $D/a$ in this probability measure: it therefore sets the
scale of the distribution of $D$ but does not alter its
functional form. The distribution of intensities is therefore
insensitive to this parameter, and we may take $a=1$.

The minima are those extrema for which $D>0$ and $T>0$.
The distribution of intensities $I\sim |D|^{-1/4}$ follows
immediately from this expression, considering only the branch
with $D>0$. The result, with arbitrary scale for the intensities, 
is given by (1.1). Numerically, the mean value of the intesities with
distribution (1.1) is found to be $\langle I\rangle=0.699..$.
The distribution (1.1) is plotted in figure 2(a).
\gap
\noindent {\sl 4.3 Relation to sea glitter}
\gap
The analysis of Gauss random functions in two dimensions
was largely motivated by the desire to achieve a statistical
understanding of the surface of the sea. It is natural to
ask which experimentally observable property of the sea
surface corresponds most closely to our distribution
of exciton transition intensities, and to compare the
two results. The corresponding property is \lq sea glitter' [14],
which is the pattern of reflections of the sun on the ocean 
surface seen from a high flying aircraft.

To simplify the dicussion we will assume that the sun is
overhead, and that the sea glitter is observed by looking
vertically downwards. If the radius of the sun subtends
an angle $\epsilon $, the observer sees a bright area 
on the sea suface due to reflected sunlight whenever the 
angle of the sea surface is smaller than ${1\over 2}\epsilon $.
There is therefore a bright spot on the sea surface in the neighbourhood
of every maximum, minimum or saddle point: each of these bright
spots is seen as an ellipse, of area $A\sim|{\rm det}(\tilde M)|^{-1}$,
where $\tilde M$ is the Hessian matrix at the stationary point.
The quantity which corresponds to the distribution of
exciton intensities is the distribution of integrated
intesities $I\sim A$ of the reflected images of the sun. The 
distribution of determinants, for all types of extrema, is
given by (4.5) and the corresponding distribution of 
$I\sim |D|^{-1}$ is
$$P(I)={\textstyle{1\over 8}}I^{-3}
\bigl[\exp(-1/2I)
+\exp(1/2I)\,{\rm erfc}({\textstyle{1\over 2}}\sqrt{3/I})\bigr]
\ .\eqno(4.6)$$
Note that, once again, the distribution is independent
of the parameter $a$: this result is therefore universal
for isotropic sea surfaces.
This distribution is plotted in figure 2(b) 
(again, the normalisation of the intensity distribution
is arbitrary, and in this case we find numerically 
$\langle I\rangle=0.433..$). This result is very different from the 
distribution of exciton intensities. The reasons for the difference are 
that all of the extrema contribute, and that there is a different
relationship between $I$ and $D={\rm det}(\tilde M)$.
\gap\gap\gap
\vfill
\eject
%
%
%%%%%%%%%%%%%%%%%%%%%%%%%%%%%%%%%%%%%%%%%%%%%%%%%%%%%%%%%%%%%%%%%%%%%
%
\noindent{\bf References}
\gap
\indentoff
\ref {1}{R. Dingle}{Festk\"orperprobleme}{15}{21}{1975}
[2] G\"obel E. O., 1988 \lq Optical Properties of Excitons in Quantum Wells', 
in {\sl Excitons in Confined Systems}, Springer: Heidelberg.
\gap
\ref {3}{C. Weisbuch, R. Dingle, A.C. Gossard and W. Weigmann}
{Solid State Commun.}{38}{709}{1981}
\ref {4}{J. Singh, K. K. Bajaj and S. Chadhuri}{Appl. Phys. Lett.}{44}{805}
{1984}
\ref {5}{J. Hegarty and M. D. Sturge}{J. Opt. Soc. Am.}{B2}{1143-54}{1985}
\ref {6}{K. P. O'Donnell and B. Henderson}{J. Luminescence}{52}{133-46}{1992}
\ref {7} {D. Gammon, E. S. Snow and D. S. Katzer} 
{Appl. Phys. Lett.}{67}{2391-3}{1995}
\ref {8} {D. Gammon, E. S. Snow, B. V. Shanabrook, D. S. Katzer and D. Park}
{Phys. Rev. Lett.}{76}{3005-8}{1996}
\ref {9}{Fang Yang, M. Wilkinson, E. J. Austin and K. P. O'Donnell}
{Phys. Rev. Lett.}{70}{323-6}{1993}
\ref {10}{M. Wilkinson, Fang Yang, E. J. Austin and K. P. O'Donnell}
{J. Phys.: Condensed Matter}{4}{8863-78}{1992}
[11] C. E. Porter (ed.), {\sl Statistical Theories of Spectra: Fluctuations},
New York: Academic Press, (1965).
\gap
\ref {12}{S. O. Rice}{Bell. Sys. Tech. J.}{24}{46-156}{1945}
\ref {13}{M. S. Longuet-Higgins}{Phil. Trans. Roy. Soc.}{A249}{321-87}{1957}
\ref {14}{C. Cox and W. Munk}{J. Opt. Soc. Am.}{44}{838}{1954}
\ref {15}{J. Hegarty, M. D. Sturge, C. Weisbuch, A. C. Gossard and W. Weigmann}
{Phys. Rev. Lett.}{49}{930-2}{1982}
[16] R. S. Knox, {\sl Theory of Excitons}, New York: Academic Press, (1963).
\gap
[17] Abramowitz M. and Stegun I. A., 1970, 
{\sl Handbook of Mathematical Functions},
Dover: New York.
\gap
\ref{18}{R. P. Stanley, J. Hegarty, R. Fischer, J. Feldmann, E. O. G\" obel,
R. D. Feldmann and R. F. Austin}{Phys. Rev. Lett.}{67}{128}{1991}
\vfill
\eject
\indentoff
{\noindent \bf Figure Captions}
\gap\gap
Figure 1. Experimental data from [15]: a) absorption spectrum, 
b) luminescence spectrum, isolated points -- resonant Rayleigh 
scattering. Theoretical curves: c) Gaussian fit to absorption
spectrum, d) corresponding distribution of heights of local
minima.
\gap\gap
Figure 2. Predicted distribution of intensities of: a) exciton
luminescence, b) sea glitter sparkles.
\vfill
\eject
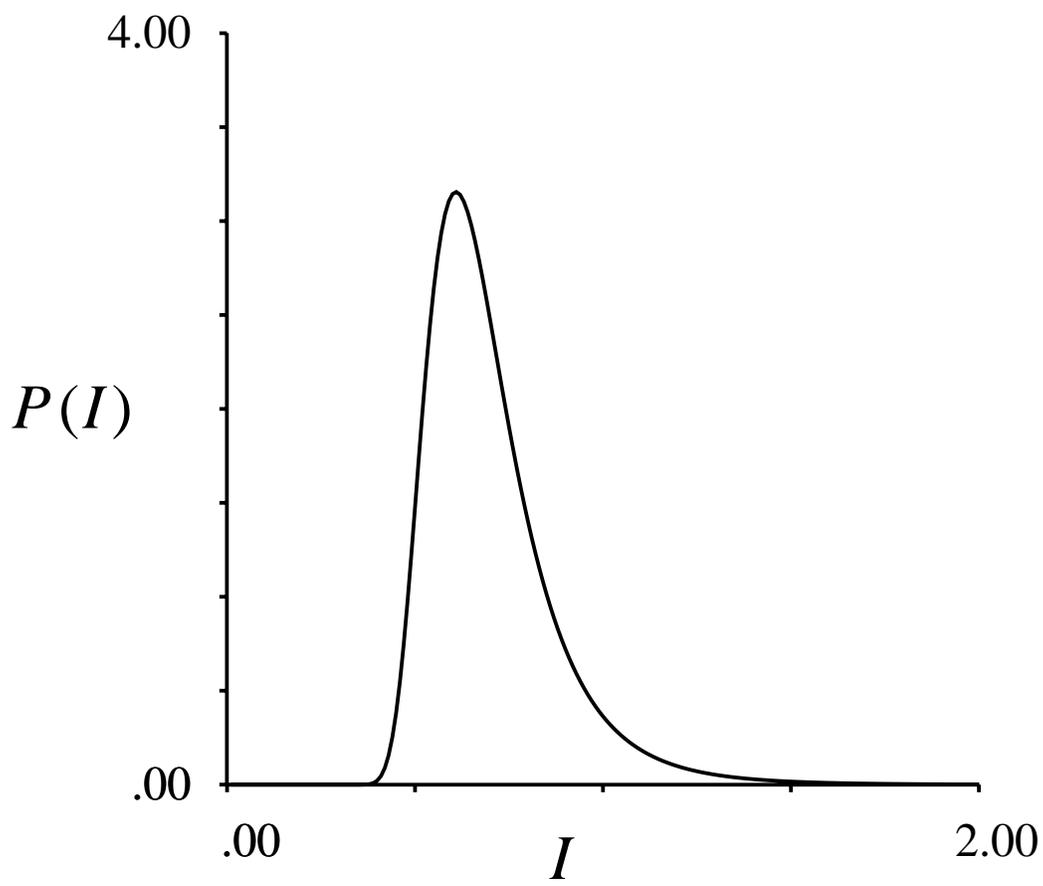
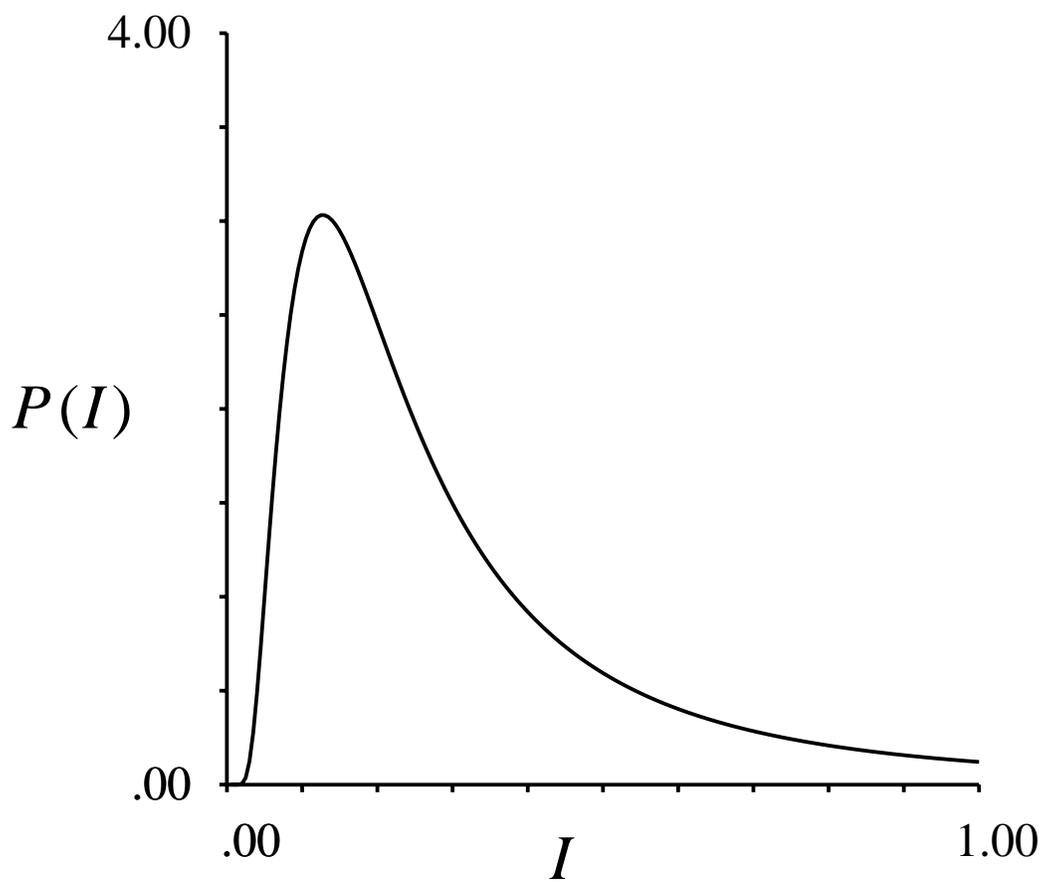
\end